\documentclass[onecolumn]{aastex62}
\usepackage{graphics,epsf}
\usepackage{amsmath}                
\usepackage{amsfonts}               
\usepackage{amssymb}                
\usepackage{epsfig}                 
\usepackage{graphicx}
\usepackage{float}
\usepackage{color}
\usepackage[para,online,flushleft]{threeparttable}

\newcommand{\yr}{{~\rm yr}}

\begin{document}

\title{Shaping planetary nebulae with jets and the grazing envelope evolution} 


\author[0000-0003-0375-8987]{Noam Soker}
\affiliation{Department of Physics, Technion, Haifa, 3200003, Israel; noa1kaplan@campus.technion.ac.il; soker@physics.technion.ac.il}
\affiliation{Guangdong Technion Israel Institute of Technology, Shantou 515069, Guangdong Province, China}

\begin{abstract}
 I argue that the high percentage of PNe that are shaped by jets show that main sequence stars in binary systems can accrete mass at a high rate from an accretion disk and launch jets. Not only this allows jets to shape PNe, but this also points to the importance of jets in other types of binary systems and in other processes. These processes include the grazing envelope evolution (GEE), the common envelope evolution (CEE), and the efficient conversion of kinetic energy to radiation in outflows. As well, the jets point to the possibility that many systems launch jets as they enter the CEE, possibly through a GEE phase. The other binary systems where jets might play significant roles include intermediate-luminosity optical transients (ILOTs), supernova impostors (including pre-explosion outbursts), post-CEE binary systems, post-GEE binary systems, and progenitors of neutron star binary systems and black hole binary systems. One of the immediate consequences is that the outflow of these systems is highly-non-spherical, including bipolar lobes, jets, and rings.
\end{abstract}


\newcommand{\orcidauthorA}{0000-0003-0375-8987} 
\section{Introduction}
 \label{sec:Intro}

The early recognition that jets play major roles in shaping planetary nebulae (PNe; e.g., \cite{Morris1987, Soker1990AJ, SahaiTrauger1998}) has received great attention and support in recent years (e.g., 
\cite{RechyGarciaetal2017, AkashiSoker2018, Balicketal2019, Derlopaetal2019, EstrellaTrujillo2019, Tafoyaetal2019, Balicketal2020, RechyGarciaetal2020}).
The binary system might launch jets before and/or after the common envelope evolution (CEE; e.g., \cite{Tocknelletal2014}).
In any case, in most cases the jets are coeval with the main nebula \citep{Guerreroetal2020}. 
The precessing of jets in many cases, e.g., as in Fleming~1 \citep{Boffinetal2012}, show that the launching mechanism is a precessing accretion disk around a compact star, like a main sequence star, a white dwarf (WD), or the core of the asymptotic giant branch (AGB) progenitor of the PN, rather than a collimation by a large-scale equatorial torus or by the AGB envelope. 
 
To form an accretion disk that might efficiently launch jets,  the accretion flow should have a large specific angular momentum. Binary interaction, mostly mass transfer, must supply the angular momentum in evolved stars. Indeed, observations and their interpretations indicate that most PNe with axisymmetrical and/or point-symmetrical morphological features are born in binary systems  (e.g., \cite{Bujarrabaletal2018, Wessonetal2018, Jones2019H, Kovarietal2019, Miszalskietal2019MNRAS487, Oroszetal2019}).  

There are some general properties in the operation of jets which I take to hold in shaping PNe (\cite{Soker2016Rev} for a review of all these properties). 
(1) The jets' velocity is about the escape speed from the object that launches the jets. For main sequence companions, the most common companions in shaping PNe, the escape speed is $v_{\rm esc} \simeq 600~{\rm km~s}^{-1}$, implying jet velocities in the general range of $v_{\rm j} \simeq (0.5-2) v_{\rm esc} \simeq 300 -1200 ~{\rm km~s}^{-1}$ .    
(2) The jets influence the gas in the ambient medium by depositing energy and momentum to the gas, i.e., by heating and/or expelling it. As such, there is less mass available for accretion onto the object that launches the jets, therefore reducing the jets' power. Namely, the jets operate in a negative feedback cycle. In particular, the jets might shut-themselves down by removing the gas available for accretion, and resume activity later on as accretion renews. This might repeat itself several times (the different mass ejection episodes will merge later to one nebula). (3) Alongside the negative feedback cycle there is a positive feedback cycle due to the jets. As jets are likely to remove energy and high entropy gas, they allow more gas to flow-in, and by that increase the mass accretion rate (e.g., \cite{Shiberetal2016}). \cite{Chamandyetal2018} remove mass and pressure via a subgrid mechanism, and  show that this `pressure release valve' mechanism, as jets are expected to act, allows high accretion rates. (4) The jets are not necessarily narrow. Jets might be wide, with a half-opening angle of almost $90^\circ$ (e.g., observations by \cite{Bollenetal2019}).

Several studies have used post-CEE binary central stars of PNe to constrain the physics and parameters of the CEE (e.g.,   \cite{DeMarco2011, Jones2020CEE}). Jones (2020; \cite{Jones2020CEE}) presents a thorough review on the usage of post-CEE PNe to learn about the CEE. Here I concentrate on the role of jets and use some properties of PNe and post-AGB binary systems to learn about the importance and outcome of the grazing envelope evolution (GEE). 
  
\section{The grazing envelope evolution (GEE)}
 \label{sec:GEE}

In the GEE (suggested in 2015 \cite{Soker2015GEE}), the jets that the secondary star launches manage to eject the envelope outside its orbit (for numerical simulations see, e.g., \cite{ShiberSoker2018, LopezCamaraetal2019}). During the GEE the secondary star grazes the envelope, including being somewhat outside the photosphere or somewhat inside it. The accretion process of gas from the giant envelope to the companion is a combination of Bondi-Hoyle-Lyttleton (BHL) accretion flow from the surrounding gas, and a Roche lobe overflow (RLOF) from the inner dense envelope. Rather than entering directly the CEE, the system performs the GEE. It might later enter a CEE or not. In some cases the secondary might get into the envelope and then exits back (e.g., \cite{Naimanetal2020}). This process might repeat itself.
   
The GEE phase has several effects and outcomes which I list as follows.
\subsection{Preventing the CEE.} 
\label{subsection:PreventingCEE}
By removing mass at a high rate the jets might prevent the CEE altogether (e.g., \cite{AbuBackeretal2018}). This process works in a negative feedback cycle, as when the orbital separation increases the accretion rate decreases and so is the effect of the jets. Tidal forces then bring the system back into contact resuming accretion at a high rate, and so on. The outcome might be a binary system with an intermediate orbital separation, $a \approx 1~{\rm AU}$. There is a group of post-AGB binary systems with such orbital separations where in some the companion is observed to launch jets (e.g., \cite{Wittetal2009, Gorlovaetal2015, Bollenetal2019}).
The GEE, therefore, might account for these post-AGB intermediate binaries (post-AGBIBs; \cite{AbuBackeretal2018}). 
\subsection{Postponing the CEE and removing mass before the CEE.}
In some cases the GEE does not prevent the CEE, but postpones it and as a result of that the system losses large amount of mass before entering the CEE. Even without launching jets, many binary systems can lose large amounts of mass before entering the CEE \citep{BearSoker2010}.  \cite{Jones2020CEE} concludes in his review that both post-CEE PNe and merging stars (luminous red novae) strongly suggest that appreciable mass transfer/loss occurs before the onset of the CEE.
\subsection{Shaping the outflow.} The interaction of the jets with the outskirts of the envelope of the giant star leads to a complicated flow structure (e.g., \cite{Shiberetal2017}). In this phase the jets mostly interact with the envelope outskirts and with the dense base of the wind, and do not expand to large distances (section \ref{sec:phases}). 
\subsection{Counteracting tidal circularization.} 
Substantially enhanced mass-loss at periastron passages during the GEE can counteract the circularization effect due to tidal interaction \citep{KashiSoker2018cir}. This might account for the non-zero eccentricity of binary systems, such as post-AGBIBs. 
 
One outcome of such a process might be a nebula that possesses departure from pure axisymmetrical structure (e.g. \cite{SokerRappaport2001} and examples therein). 

\subsection{Experiencing ILOT events.} The jets that the companion launches collide with envelope gas, and if the jets penetrate out they might collide with the shell that the high mass loss rate from the giant star forms. The collision of the jets with the envelope (e.g., \cite{Soker2016GEEILOT}) and/or with the slow shell (e.g., \cite{Soker2020jetILOT})
converts kinetic energy to thermal energy. If the photon diffusion time out is not much longer than the expansion time of the gas (which is the adiabatic cooling time), then a large fraction of the thermal energy is carried out by radiation. In a short GEE episode, less than few years, the observation will be of an intermediate luminous optical transient, namely, and ILOT (which might be brightest in the IR if efficient dust formation takes place).  
\subsection{Forming type IIb supernovae (SNe~IIb).} SNe~IIb are core collapse supernovae with little hydrogen mass at explosion ($M_{\rm H,ex} \simeq 0.01-1 M_\odot$). The enhanced mass loss due to jets in a binary system of a giant progenitor of core collapse supernova and a main sequence companion that experiences the GEE, might leave little hydrogen mass at explosion, hence forming a SN~IIb progenitor \citep{Naimanetal2020}. This is only one of several evolutionary channels that might lead to SNe~IIb. 

\section{Evolutionary phases of jets' launching}
 \label{sec:phases}
 
For the purpose of this study I consider five evolutionary phases where the companion to a giant star might launch jets.  
I summarise these in two tables. In Table \ref{Tab:Table1} I list three phases where in most cases the jets do not expand to large distances. In Table \ref{Tab:Table2} I list the two  phases where in most cases the jets might expand to large distances and form bipolar lobes and/or other morphological features along and near the symmetry axis. 
\begin{table}
\caption{Evolutionary phases with choked jets}
\centering
\begin{tabular}{p{5.8cm}p{5.8cm}p{5.8cm}}
\hline
\textbf{1. Jets in wind acceleration zone}	& \textbf{3. Jets during the GEE}	& \textbf{4. Jets during the CEE} \\
\hline
Bondi--Hoyle--Lyttleton (BHL) accretion from the slow wind. & BHL + Roche lobe overflow (RLOF) accretion from envelope outskirts. & BHL accretion inside the giant envelope. \\
\includegraphics[trim= 0cm 5.5cm 0cm 2.5cm ,clip=true,width=0.35\textwidth]{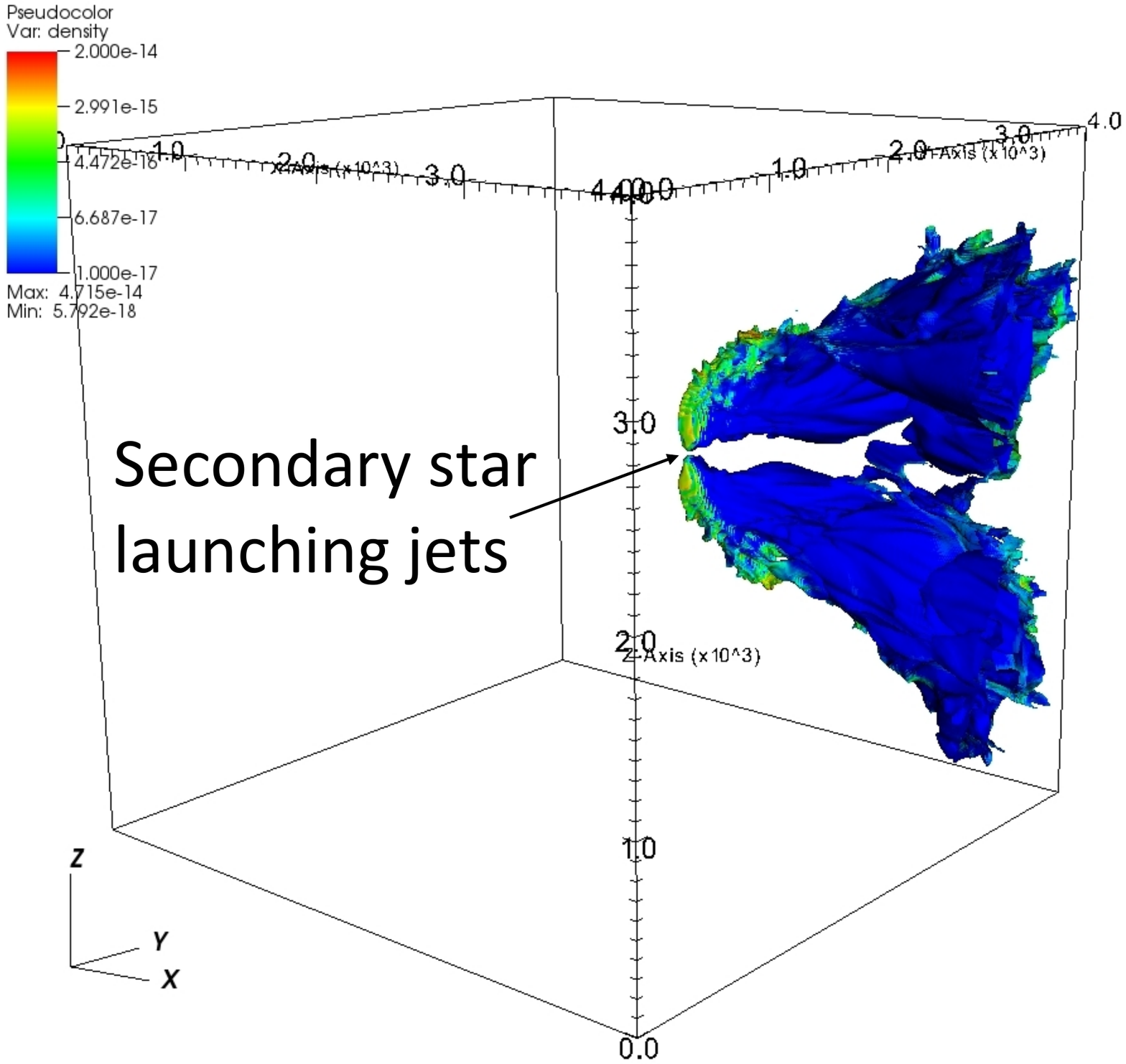}		 & 
\includegraphics[trim= 1.0cm 8.5cm 0cm 2.5cm ,clip=true,width=0.35\textwidth]{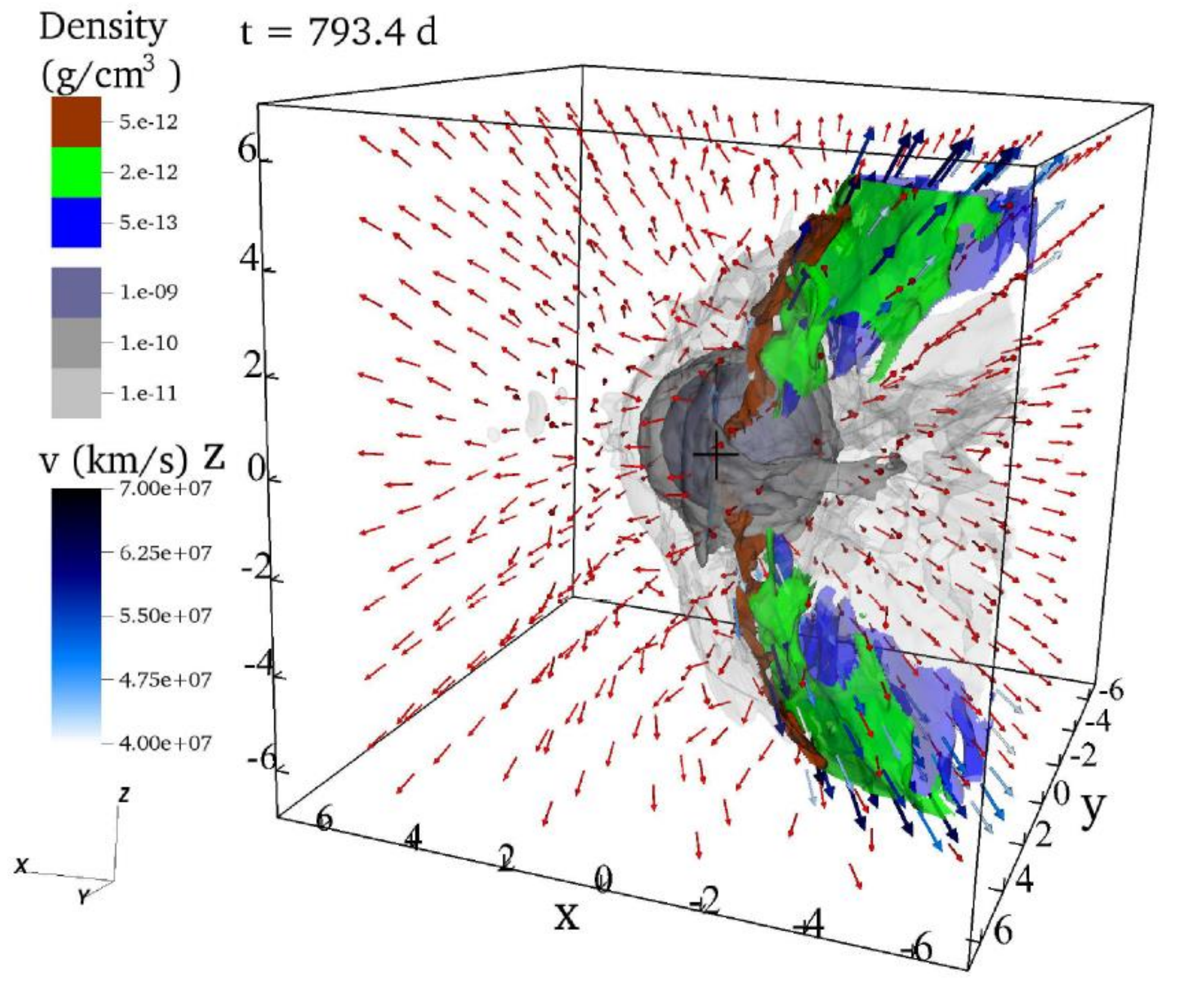}		& 
\includegraphics[trim= 0cm 3.8cm 0cm 2.0cm ,clip=true,width=0.32\textwidth]{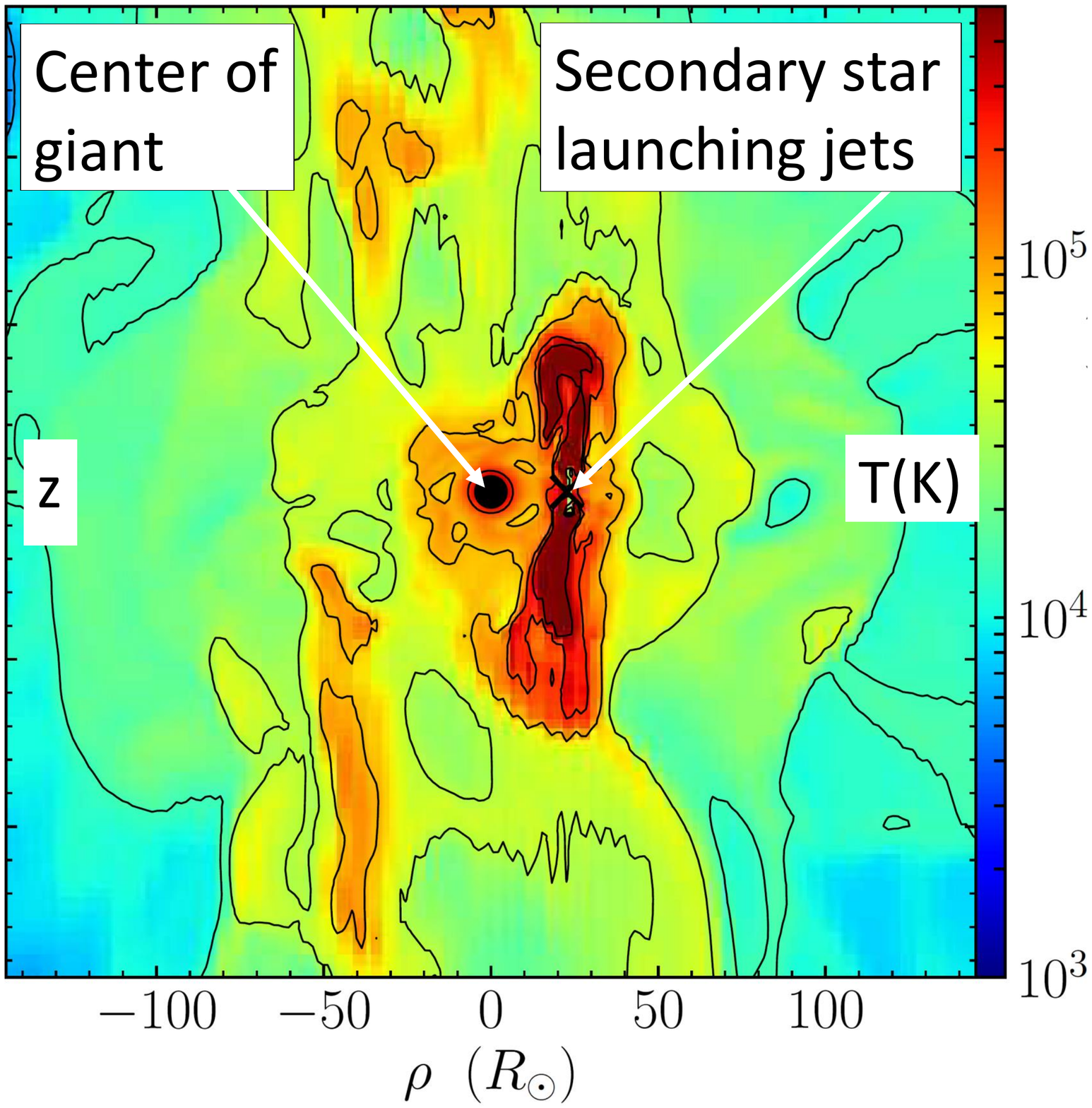}			\\
 & & \\
Density-surface 3D maps of the gas that originated in the jets and mixed with the wind. 
Colours are density surfaces according to the colour bars on the left from $10^{-17}$ (blue) to $2 \times 10^{-14} {\rm g~ cm}^{-1}$ (red) (from \cite{Hilleletal2020}). 
& 
The AGB is at the center. The secondary star (marked by``$+$'') is moving to the left in the figure.
The red-green-blue colors show gas originated in the jets and moving with velocities $>400~\rm{km~s^{-1}}$. The gray marks slow AGB gas. Axes from $-7\times 10^{13}$ to $7\times 10^{13} {\rm cm}$.
(For more details see original study \cite{Shiberetal2017}).
& 
A temperature map (bar in K) in the meridional plane of a 3D CEE simulation. The secondary launches two opposite jets, in the $+z$ and $-z$ directions, as it orbits inside the envelope of a giant star. The high temperature (red) of the shocked jets gas show the jets do not break out from the envelope, i.e., they are choked jets (from \cite{Shiberetal2019}).
\\
\hline
\end{tabular}
\label{Tab:Table1}
\end{table}
\begin{table}
\caption{Evolutionary phases with expanding jets}
\centering
\begin{tabular}{p{8.5cm}p{8.5cm}}
\hline
\textbf{2. Jets from pre-CEE RLOF accretion}	& \textbf{5. Jets during the post-CEE}	\\
\hline
Accretion onto the secondary star by RLOF from the giant envelope (+ wind-RLOF accretion from wind acceleration zone). 
& The secondary star accretes from the post-CEE circumbinary disk (the remnant of the envelope) and launches jets. \\ 
\includegraphics[trim= 2.5cm 10.9cm 0cm 2.0cm ,clip=true,width=0.48\textwidth]{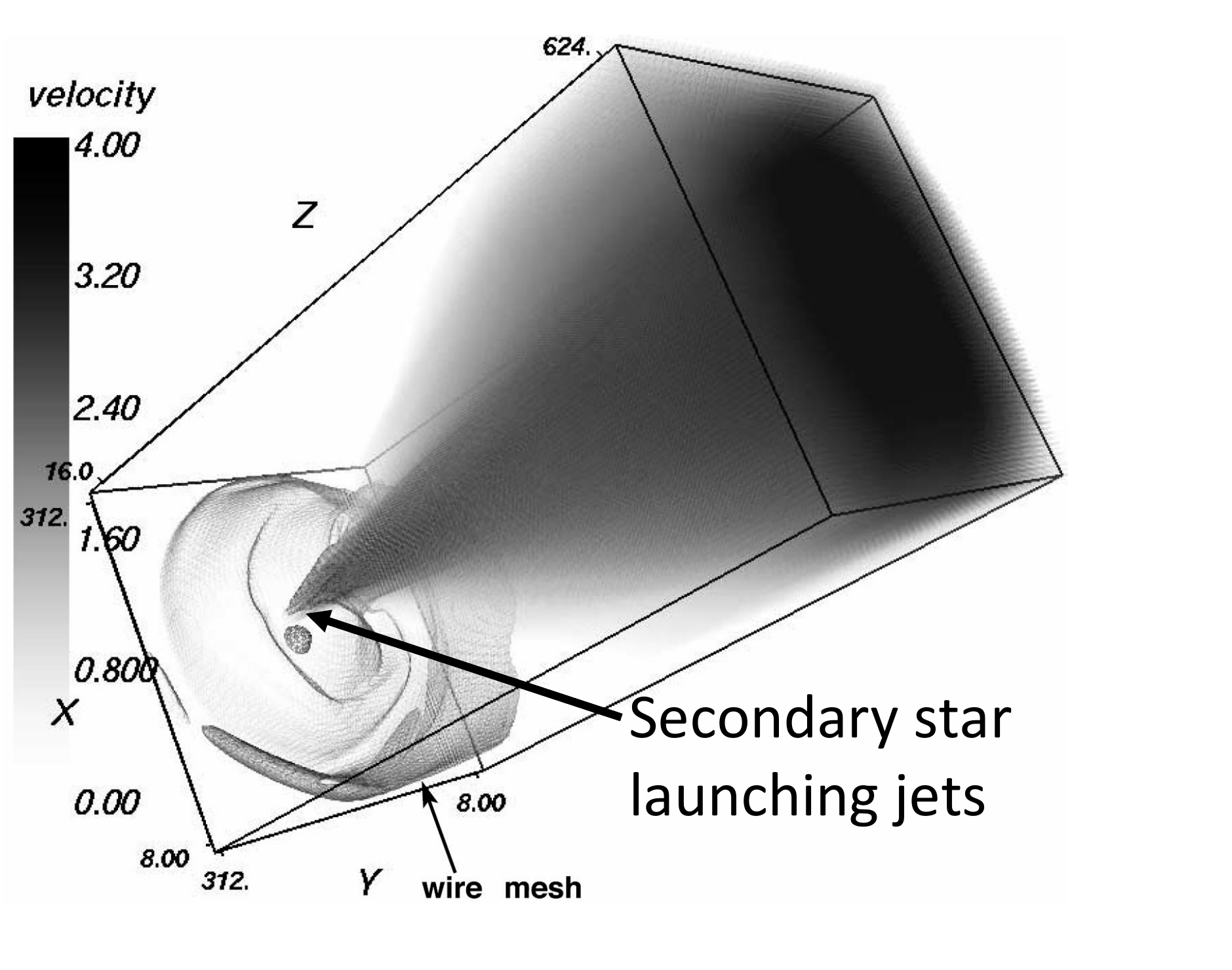}	  
&
\includegraphics[trim= 7.6cm 16.5cm 3.0cm 4.0cm ,clip=true,width=0.45\textwidth]{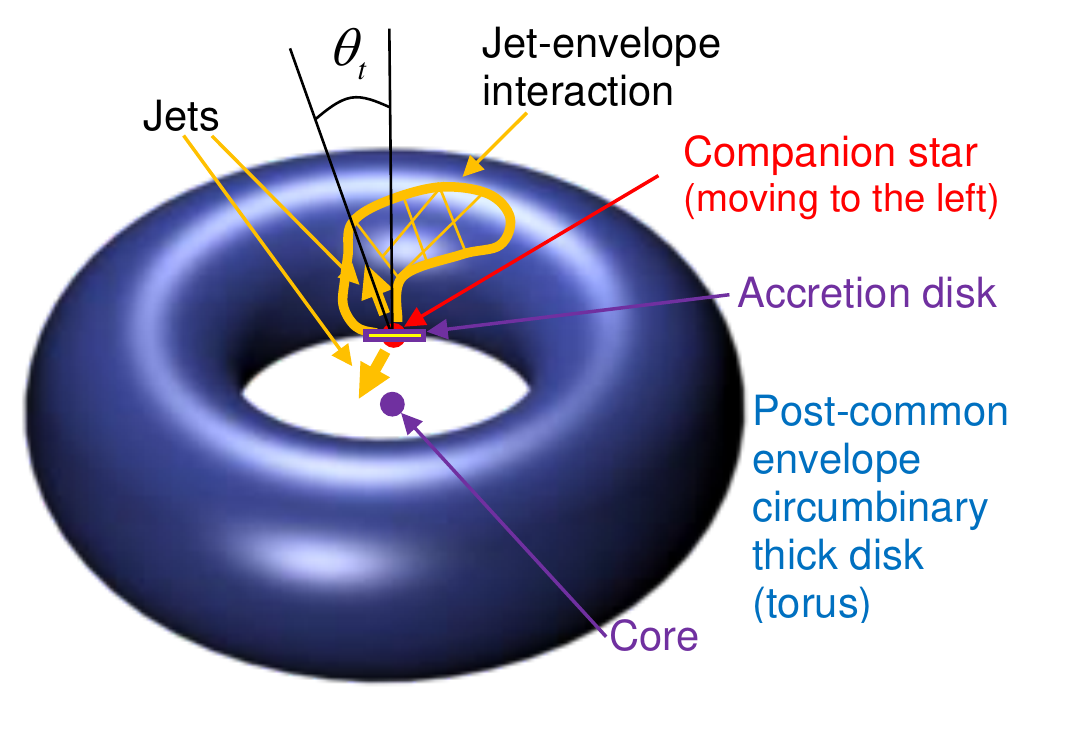}			\\
Velocity map in 3D together with density isosurface (\textit{wire mesh}) from the 3D hydro simulations of \cite{GarciaArredondoFrank2004}. Axes are in pixels. Jets manage to entrain AGB wind and expand out.  
&
A schematic drawing (not to scale) of the last phase of the CEE. The orange hatched region depicts the hot bubble formed by the interaction of one jet with the envelope (opposite bubble not shown). At later times the jets might expand almost freely (from \cite{Soker2019Termination}). 
\\
\hline
\end{tabular}
\label{Tab:Table2}
\end{table}
  
\textit{1. Jets in wind acceleration zone} (first column of Table \ref{Tab:Table1}). The secondary star can remove mass before it grazes the envelope itself, if it launches jets in the acceleration zone of the dense wind of the giant star, as numerical simulations show \cite{Hilleletal2020}. When the accretion rate is mostly due to BHL accretion flow, the jets from main sequence stars are not strong enough to penetrate through the wind from which the secondary star accretes mass \citep{Hilleletal2020}. 

\textit{2. Jets from pre-CEE RLOF accretion} (first column of Table \ref{Tab:Table2}). When the secondary is closer to the surface, the accretion is mainly due to RLOF from the envelope. Most of the accretion flow is equatorial and it is much larger than that of BHL accretion from the wind through which the secondary orbits the giant star. In such a case the jets are likely to penetrate through the wind and expand to large distances, therefore shaping the slow outflow to form a bipolar morphology with one or more pairs of lobes and/or other point symmetric morphologies (e.g., \cite{GarciaArredondoFrank2004}). This case might occur also at a somewhat earlier phase when the dense acceleration zone of the wind overflows its Roche lobe and transfers mass at a high rate to the secondary star via an equatorial flow (for this wind-RLOF see, e.g., \cite{Harpazetal1997, MohamedPodsiadlowski2007}). 

\textit{3. Jets during the GEE} (second column of Table \ref{Tab:Table1}). This phase was the subject of section \ref{sec:GEE}. The jets strongly interact with the envelope, and in most cases do not expand to large distances. The situation might be different in an eccentric orbit (as might have been the case during the Great Eruption of Eta Carinae, e.g., \cite{KashiSoker2010}), when the secondary star accretes mass when it grazes the giant envelope near periastron, and the jets' launching episode lasts to later times when the secondary star is near apastron. 

\textit{4. Jets during the CEE} (third column of Table \ref{Tab:Table1}). Inside the envelope the density is too large for jets that a main sequence star launches to expand out (e.g., \cite{Shiberetal2019}). The jets might play a crucial role in helping mass removal. 

\textit{5. Jets during the post-CEE} (second column of Table \ref{Tab:Table2}). In a recent paper \citep{Soker2019Termination} I discussed the possibility that the secondary star accretes mass from a circumbinary disk \citep{KashiSoker2011, ChenPodsiadlowski2017} at the termination of the CEE. As a result of that the secondary star might launch jets that have variable directions and intensities. These jets shape the nebula. Note that in this mechanism the secondary star (or maybe even the core of the former AGB star) launches the jets. This should be distinguish from a process where the highly deformed envelope collimates a bipolar outflow (e.g., \cite{Soker1992, Zouetal2020}). 

\section{Supporting observations to the GEE}
 \label{sec:Properties}
 
In this section I list some observations that support to some extend the occurrence of the GEE. There is no observation (yet) that directly shows this phase to take place, but I take the observations I list in Table \ref{Tab:supports} to indirectly support the GEE. These observations are of two kinds, those that show the importance of jets (first four items of Table \ref{Tab:supports}), and those that raise some puzzles that the GEE might solve (last two rows of Table \ref{Tab:supports}).  
\begin{table}
\caption{Observational indirect support to the importance of the GEE. }
\centering
\begin{tabular}{p{4.3cm}p{2.5cm}p{7.4cm}}
\hline
\textbf{Property}	& \textbf{Example}	& \textbf{Possible implications}\\
\hline
Jets launching $\approx 10^3-10^4 \yr$ before main nebular ejection.  &  ETHOS~1 {\citep{Miszalskietal2011}}	& A strong binary interaction which involves mass transfer and jet launching takes place shortly before the CEE. Might be phase 2 (left column) of Table \ref{Tab:Table2}. \\
\hline
Jets launching $\approx 10^3-10^4 \yr$ after main nebular ejection.  & Hb4 {\citep{Derlopaetal2019}}			& A strong binary interaction which involves mass transfer and jet launching takes place shortly after the CEE. Might be mass accretion from a circumbinary disk, as phase 5 in Table \ref{Tab:Table2}. \\
\hline
High-momentum jets in pre-PNe (or in PNe).        & M1-92 {\citep{Bujarrabaletal2001}} & Main sequence stars can accrete mass at high rates, $\dot M > 10^{-4} M_\odot \yr^{-1}$, and launch jets when  approaching, inside, or just exiting a CEE {\citep{BlackmanLucchini2014}}.          \\
\hline
Post-AGBIBs with jets and  a circumbinary disk.  & IRAS~19135+3937 {\citep{Bollenetal2019}}	& (1) Main sequence companion outside an AGB star can launch jets that shape the outflow. (2) Accretion might take place from a circumbinary disk {\citep{Bollenetal2019}}. \\
\hline
Low-mass nebulae in some post-CEE PNe  {\citep{SantanderGarciaetal2019}}. &  Abell~63 {\citep{Corradietal2015}}     & Significant pre-CEE mass loss; possibly in a GEE  {\citep{SantanderGarciaetal2019}} .  \\
\hline
PN central binaries with orbital periods of $P_{\rm orb} >{\rm several~day}$. & NGC~2346 {\citep{Brownetal2019}} & Significant pre-CEE mass loss; possibly in a GEE {\citep{Soker2015GEE}}. \\
\hline
\end{tabular}
\label{Tab:supports}
\end{table}

\textit{Jets launching $\approx 10^3-10^4 \yr$ before or after main nebular ejection.}
Observations suggest that in some cases PN progenitors launch jets shortly before, during, or after they eject the main nebulae (e.g., \cite{Huggins2007, Tocknelletal2014, JonesBoffin2017, Guerreroetal2020}).
In the second and third rows of Table \ref{Tab:supports} I list two examples and summarise the possible implications of these observations.  

\textit{High momentum jets in PNe.}
\cite{BlackmanLucchini2014} study 19 pre-PNe (taken from \cite{Bujarrabaletal2001, Sahaietal2008}) and constrain the minimum mass accretion rate to account for the high momenta of the bipolar outflows (jets). They consider main sequence and white dwarf (WD) accretors that launch the jets. They conclude that accretion from the wind is not sufficient to explain the momentum in the jets. Main sequence companions accreting via a RLOF or inside a common envelope can supply the required momentum. As for WD accretors, in most cases that they study the accretion rate is super-Eddington. As I discuss in section \ref{sec:Secondary}, another problem with WDs accreting at a high rate is that they ignite nuclear reactions and inflate an envelope. 
Overall, the conclusion from the work of \cite{BlackmanLucchini2014} is that main sequence stars can accrete at high rates, $\dot M > 10^{-4} M_\odot \yr^{-1}$, just before, during, and/or just after the CEE (third item in Table \ref{Tab:supports}). 

\textit{Post-AGBIBs with jets and a circumbinary disk}. Traditional CEE calculations suggest that either the binary system enters a CEE and the final orbital separation is much smaller than the giant radius ($\approx 1 {\rm AU}$), or that mass loss causes the orbital separation to increase much above the giant radius. However, there is a group of post-AGB intermediate binaries (post-AGBIBs) with orbital separations of $\approx 1 {\rm AU}$. I already mentioned this group in section \ref{subsection:PreventingCEE}, where I suggested that the GEE might prevent the  formation of a CEE, such that the orbital separation stays at about the giant radius. The point I emphasise here is that observations suggest that the main sequence companion in most post-AGBIBs launches jets (e.g., \cite{Wittetal2009, Gorlovaetal2015, Bollenetal2019}). I list the implications in the fourth item of Table \ref{Tab:supports}. The second implication, that the companion accrete mass from a circumbinary disk and launch jets, is directly relevant to post-CEE binaries, as indicated in the second column of Table \ref{Tab:Table2}.

\textit{Low-mass nebulae in some post-CEE PNe.} There are several hints that a large fraction of systems enter the CEE after they have lost a significant amount of mass (e.g., \cite{Jones2020CEE} for a recent paper). \cite{SantanderGarciaetal2019} as another example, claim for lower nebular mass in post-CEE PNe. This again suggests a substantial pre-CEE mass loss, which the GEE might account for \citep{SantanderGarciaetal2019}. Many episodes of GEE might help in removing large amounts of mass. Another outcome of significant pre-CEE mass loss might be a large post-CEE orbital separation \citep{Soker2015GEE}, such as in the central binary system of NGC~2346.   

\textit{PN central binaries with orbital periods of $P_{\rm orb} >{\rm several~days}$.} Central binary systems of PNe with orbital separations below $\approx 100 R_\odot$ have experienced the CEE (rather than only the GEE for larger orbital separations). If the orbital separation is above $\approx 10 R_\odot$, corresponding to an orbital period of $P_{\rm orb} > 4~{\rm days}$, the possible implication is that the envelope mass at the beginning of the CEE was small, such that the binary system ejected the entire envelope when the orbital separation was relatively large. The GEE might be a way to remove mass in the pre-CEE phase. 

\section{The mass-accreting secondary star}
 \label{sec:Secondary}
 In earlier sections I referred mainly (but not only) to main sequence secondary stars.  Here I list also other possibilities. I list by increasing maximum jets' power.  
  
\subsection{Planets as secondary objects.}
\label{subsec:planets}
In cases of low mass PN progenitors where the final envelope mass is low, $M_{\rm env} \simeq 0.1 - 0.5 M_\odot$, a planet can play a role in shaping the outflow. This scenario is made more likely if the mass loss rate of AGB stars that do not acquire any angular momentum from a companion (so called Jsolated stars) is lower than what traditional formulae give (e.g., \cite{SabachSoker2018}). 

Planets might accrete small amounts of mass, and at best launch jets at velocities of $\approx 100~{\rm km~s}^{-1}$. As planets have no strong convection that mixes accreted mass inward, I take the mass that a planet of mass $M_{\rm p}$ can accrete before it expands to be below about $M_{\rm acc, p} \simeq 0.01- 0.1 M_{\rm p}$. The mass in each jet might be about 10 percent of that mass, 
$M_{\rm jet,p} \approx 0.001-0.01 M_{\rm p}$. Scaling this relation with a massive planet (about 10 times the mass of Jupiter) gives 
$M_{\rm jet} \approx 10^{-5} -10^{-4} (M_{\rm p}/0.01M_\odot) M_\odot$. 

Interestingly, there are two opposite knots along the symmetry axis of some elliptical PNe, termed `ansae' (or FLIERS), whose typical velocities and masses are $\approx {\rm few} \times 10 -200 ~{\rm km~s}^{-1}$ and $\approx 10^{-5}-10^{-4} M_\odot$, respectively \citep{Balicketal1994}. There is a need for more studies to examine whether massive planets and brown dwarfs might launch jets in shaping PNe, and whether in some cases these jets might form ansae (FLIERS). 

There is an indirect way for a planet (or a brown dwarf) to form jets. If the core of the AGB star tidally destroys the planet to form an accretion disk at the termination of the CEE, then the core might launch jets made from the destroyed planet material (e.g. \cite{Soker1996, ReyesRuizLopez1999, Blackmanetal2001, NordhausBlackman2006, Guidarellietal2019}). These jets, thought, will be much faster than the observed velocities of ansae.

\subsection{WD secondary stars.}
\label{subsec:WDs}
In \cite{Soker2004NewA} I discuss accreting WDs and main sequence stars during the CEE (note that here the WD is the secondary star, and the mass-donor giant is the primary star). 
The potential well of a WD is about two orders of magnitude deeper than that of a main sequence star. However, the mass accretion rate of a WD is limited. This is because at an accretion rate of $\dot M_{\rm WD} > \approx 10^{-6} M_\odot \yr^{-1}$ nuclear reactions on the surface of the WD inflate an envelope (e.g., \cite{Hachisuetal1999}). This upper limit is about the lower accretion rate that \cite{BlackmanLucchini2014} require to explain high-momenta bipolar outflows in pre-PNe. 

This addition of nuclear energy is about doubling the luminosity from the core of the AGB star. It is not clear whether the WD can launch jets at accretion rates above that value. Still, at lower values of $\dot M_{\rm WD} < {\rm few} \times 10^{-7} M_\odot \yr^{-1}$, the WD might launch jets at terminal velocities of $v_{\rm j,WD} \approx {\rm few} \times 1000~{\rm km~s}^{-1}$. 
Overall, WDs might launch jets at the different stages that I summarised in section \ref{sec:phases} and play a role in the different phases. I expect jets from WDs not to reach the highest powers of jets from main sequence stars. This question deserves further study. 

\subsection{Main sequence secondary stars (low masses and high masses).}
\label{subsec:MSs}

The supporting observations that I discussed in section \ref{sec:Properties} are mainly for main sequence secondary stars in low mass systems. Namely, the secondary stars are mainly in the mass range $\simeq 0.3-1 M_\odot$, and observations suggest that these secondary stars can accrete at a high rate and launch jets. These are main sequence stars with convective envelopes. They are more likely to accrete mass without expanding much. The removal of angular momentum, energy, and high entropy gas by the jets themselves \citep{Shiberetal2016}, as in the `pressure release valve' mechanism \citep{Chamandyetal2018}, helps in allowing high mass accretion. There is no question that low-mass main sequence stars can launch relatively energetic jets. 
 
I emphasise that observations (e.g.,  \cite{Jonesetal2015Exp}) do show that post-CEE main sequence stars have larger radii than their radii on the main sequence, as expected for such stars that accrete mass (e.g., \cite{PrialnikLivio1985}). The point is that the radii of these post-CEE main sequence stars increase by a factor of only $\approx 2$. Namely, the gravitational potential of the secondary star does not change much, and the escape velocity is still larger than $\approx 300 ~{\rm km~s^{-1}}$. Namely, jets that these stars launch even after they expand are still relatively fast, and they might be sufficiently energetic to shape the nebula and expel mass from the envelope. 

There is the question of whether more massive stars that have radiative envelope can accrete mass and launch jets. I think that the bipolar nebula of Eta Carinae, called Homunculus, that was formed during the Great Eruption in the nineteenth century is an evidence that the answer is yes. There are indications and suggestions (e.g., \cite{AkashiKashi2020} and references there in) that the Homunculus was shaped by jets. As \cite{AkashiKashi2020} show, jets that the secondary star launched might as well account for the very fast outflow in the Great Eruption ($v > 10^4 ~{\rm km~s}^{-1})$.
  
\subsection{Neutron stars and black holes.}
\label{subsec:NSs}

In principle, a neutron star (NS) and a black hole (BH) can be the secondary star in each of the five jet-launching phases that I discussed in section \ref{sec:phases}. In particular, during the GEE and CEE neutrino cooling allows very high mass accretion rates. The giant primary star is likely to be a massive progenitor of a core collapse supernova with a mass above $\approx 8 M_\odot$.  
The role of jets that a NS and a BH launch during the CEE (and possibly in an earlier GEE phase) must be considered in the study of the formation of NS-NS, NS-BH, or BH-BH close binary systems.  
 
The collision of jets that main sequence stars (and possibly WDs) launch with the envelope and slow outflow can lead to ILOTs, i.e., transient events that have luminosities below the typical luminosity of supernovae. 
On the other hand, the collision of the very energetic jets that NSs and BHs launch during the CEE can convert huge amounts of energy to thermal energy, part of which is radiated away, possibly as an event as bright as a typical supernova or more.  
If the NS (or BH) spirals-in all the way to the core, it destroys the core and might accrete a mass of $M_{\rm acc, NS} \approx 0.1-1 M_\odot$ within minutes \citep{GrichenerSoekr2019}. The outcome might be a very bright event, much brighter even than typical supernovae (e.g., \cite{Sokeretal2019CEJSN}). This event is termed a \textit{common envelope jets supernova} (CEJSN). One plausible outcome might be the nucleosynthesis of r-process elements in the jets \citep{GrichenerSoekr2019}.

If the NS or BH accrete mass only from the relatively low density envelope, and then gets out without completely destructing the giant star, the event is termed CEJSN impostor \citep{Gilkisetal2019}.


\section{Summary}

The observations that jets that binary systems launch shape many PNe (section \ref{sec:Intro}), before, during, and/or after the ejection of the main nebula (section \ref{sec:phases}), suggest that the companion might launch jets also when it enters the envelope and when inside the envelope (section \ref{sec:Properties}). Therefore, when studying the CEE it is mandatory to consider jets, in particular when the companion is a NS or a BH (section \ref{subsec:NSs}). These observations support the idea of the GEE (section \ref{sec:Properties}), where the jets manage to remove the envelope outside the orbit of the secondary star (section \ref{sec:GEE}). The secondary stars might be planets (or brown dwarfs), main sequence stars, WDs, NSs, and BHs (section \ref{sec:Secondary}). This study was mainly on main sequence secondary stars. 

Finally, $\approx 10 \%$ of the systems that experience the CEE might be close triple systems. This might lead to the formation of `messy nebulae', i.e., nebulae that lack any clear symmetry, i.e., by launching inclined jets \citep{Schreieretal2019}.

\acknowledgments{I thank Bruce Balick for very helpful comments. This research was supported by a grant from the Israel Science Foundation and a grant from the Asher Space Research Fund at the Technion.}


\begin{thebibliography}{999}

\bibitem[Abu-Backer et al.(2018)]{AbuBackeretal2018} Abu-Backer, A., Gilkis, A., \& Soker, N.\ \textbf{2018}, {\em ApJ}, 861, 136

\bibitem[Akashi \& Kashi(2020)]{AkashiKashi2020} Akashi, M., \& Kashi, A.\ \textbf{2020}, preprint 

\bibitem[Akashi \& Soker(2018)]{AkashiSoker2018} Akashi, M., \& Soker, N.\ \textbf{2018}, {\em MNRAS}, 481, 2754

\bibitem[Balick et al.(2019)]{Balicketal2019} Balick, B., Frank, A., \& Liu, B.\ \textbf{2019}, {\em ApJ}, 877, 30

\bibitem[Balick et al.(2020)]{Balicketal2020} Balick, B., Frank, A., \& Liu, B.\ \textbf{2020}, {\em ApJ}, 889, 13

\bibitem[Balick et al.(1994)]{Balicketal1994} Balick, B., Perinotto, M., Maccioni, A., Terzian, Y., \& Hajian, A.,\ \textbf{1994}, {\em ApJ}, 424, 800

\bibitem[Bear \& Soker(2010)]{BearSoker2010} Bear, E., \& Soker, N.\ \textbf{2010}, {\em New Astronomy}, 15, 483

\bibitem[Blackman \& Lucchini(2014)]{BlackmanLucchini2014} Blackman, E.~G., \& Lucchini, S.\ \textbf{2014}, {\em MNRAS}, 440, L16

\bibitem[Blackman et al.(2001)]{Blackmanetal2001} Blackman, E.~G., Frank, A., \& Welch, C.\ \textbf{2001}, {\em ApJ}, 546, 288

\bibitem[Boffin et al.(2012)]{Boffinetal2012} Boffin, H.~M.~J., Miszalski, B., Rauch, T.,  Jones, D., Corradi, R.~L.~M., Napiwotzki, R., Day-Jones, A.~C., \& K\"oppen, J.\ \textbf{2012}, {\em Science}, 338, 773

\bibitem[Bollen et al.(2019)]{Bollenetal2019} Bollen, D., Kamath, D., Van Winckel, H., et al.\ \textbf{2019}, {\em A\&A}, 631, A53

\bibitem[Brown et al.(2019)]{Brownetal2019} Brown, A.~J., Jones, D., Boffin, H.~M.~J., \& Van Winckel H.\ \textbf{2019}, {\em MNRAS}, 482, 4951


\bibitem[Bujarrabal et al.(2018)]{Bujarrabaletal2018} Bujarrabal V., Castro-Carrizo A., Van Winckel H., Alcolea J., S{\'a}nchez Contreras C., Santander-Garc{\'\i}a M., Hillen M., \textbf{2018}, {\em A\&A}, 614, A58


\bibitem[Bujarrabal et al.(2001)]{Bujarrabaletal2001} Bujarrabal, V., Castro-Carrizo, A., Alcolea, J., \& S{\'a}nchez Contreras C.\ \textbf{2001}, {\em A\&A}, 377, 868

\bibitem[Chamandy et al.(2018)]{Chamandyetal2018} Chamandy, L., Frank, A., Blackman, E.~G., et al.\ \textbf{2018}, {\em MNRAS}, 480, 1898
 
\bibitem[Chen \& Podsiadlowski(2017)]{ChenPodsiadlowski2017} Chen, W.-C., \& Podsiadlowski, P.\ \textbf{2017}, {\em ApJL}, 837, L19

\bibitem[Corradi et al.(2015)]{Corradietal2015} Corradi, R.~L.~M., Garc{\'\i}a-Rojas, J., Jones, D., \& Rodr{\'\i}guez-Gil, P.\ \textbf{2015}, {\em ApJ}, 803, 99


\bibitem[De Marco et al.(2011)]{DeMarco2011} De Marco, O., Passy, J.-C., Moe, M., Herwig, F., Mac Low, M.-M., \& Paxton B., \textbf{2011}, {\em MNRAS}, 411, 2277

\bibitem[Derlopa et al.(2019)]{Derlopaetal2019} Derlopa S., Akras S., Boumis P., Steffen W., \textbf{2019}, {\em MNRAS}, 484, 3746

\bibitem[Estrella-Trujillo et al.(2019)]{EstrellaTrujillo2019}  Estrella-Trujillo D., Hern{\'a}ndez-Mart{\'\i}nez L., Vel{\'a}zquez P.~F., Esquivel A., Raga A.~C., \textbf{2019}, {\em ApJ}, 876, 29

\bibitem[Garc{\'\i}a-Arredondo \& Frank(2004)]{GarciaArredondoFrank2004} Garc{\'\i}a-Arredondo \& Frank Garc{\'\i}a-Arredondo, F., \& Frank, A.\ \textbf{2004}, {\em ApJ}, 600, 992

\bibitem[Gilkis et al.(2019)]{Gilkisetal2019} Gilkis, A., Soker, N., \& Kashi, A.\ \textbf{2019}, {\em MNRAS}, 482, 4233

\bibitem[Gorlova et al.(2015)]{Gorlovaetal2015} Gorlova, N., Van Winckel, H., Ikonnikova, N.~P., Burlak, M.~A., Komissarova, G.~V., Jorissen, A., Gielen, C., Debosscher, J., \& Degroote, P. \textbf{2015}\, {\em MNRAS}, 451, 2462

\bibitem[Grichener \& Soker(2019)]{GrichenerSoekr2019} Grichener, A., \& Soker, N.\ \textbf{2019}, {\em ApJ}, 878, 24

\bibitem[Guerrero et al.(2020)]{Guerreroetal2020} Guerrero, M.~A., Rechy-Garcia, J.~S., \& Ortiz, R.\ \textbf{2020}, arXiv:1912.07754


\bibitem[Guidarelli et al.(2019)]{Guidarellietal2019} Guidarelli, G., Nordhaus, J., Chamandy, L., et al.\ \textbf{2019}, {\em MNRAS}, 490, 1179

\bibitem[Hachisu et al.(1999)]{Hachisuetal1999} Hachisu, I., Kato, M., \& Nomoto, K.\ \textbf{1999}, {\em ApJ}, 522, 487

\bibitem[Harpaz et al.(1997)]{Harpazetal1997} Harpaz, A., Rappaport, S., \& Soker, N.\ \textbf{1997}, {\em ApJ}, 487, 809 

\bibitem[Hillel et al.(2020)]{Hilleletal2020} Hillel, S., Schreier, R., \& Soker, N.\ \textbf{2020}, arXiv e-prints, arXiv:1912.04662

\bibitem[Huggins(2007)]{Huggins2007} Huggins, P.~J.\ \textbf{2007}, {\em ApJ}, 663, 342

\bibitem[Jones(2019)]{Jones2019H} Jones, D.\ \textbf{2019}, Highlights on Spanish Astrophysics X, 340

\bibitem[Jones(2020)]{Jones2020CEE} Jones, D.\ \textbf{2020}, arXiv e-prints, arXiv:2001.03337

\bibitem[Jones \& Boffin(2017)]{JonesBoffin2017} Jones, D., \& Boffin, H.~M.~J.\ \textbf{2017}, {\em Nature Astronomy}, 1, 0117

\bibitem[Jones et al.(2015)]{Jonesetal2015Exp} Jones, D., Boffin, H.~M.~J., Rodr{\'\i}guez-Gil, P., Wesson, R., Corradi, R.~L.~M., Miszalski, B., \& Mohamed S.\ \textbf{2015}, {\em A\&A} 580, A19

\bibitem[Kashi \& Soker(2010)]{KashiSoker2010} Kashi, A., \& Soker, N.\ \textbf{2010}, {\em ApJ}, 723, 602

\bibitem[Kashi \& Soker(2011)]{KashiSoker2011} Kashi, A., \& Soker, N.\ \textbf{2011}, {\em MNRAS}, 417, 1466

\bibitem[Kashi \& Soker(2018)]{KashiSoker2018cir} Kashi, A., \& Soker, N.\ \textbf{2018}, {\em MNRAS}, 480, 3195

\bibitem[K{\H{o}}v{\'a}ri et al.(2019)]{Kovarietal2019} K{\H{o}}v{\'a}ri, Z., Strassmeier, K.~G., Ol{\'a}h, K., et al.\ \textbf{2019}, {\em A\&A}, 624, A83

\bibitem[L{\'o}pez-C{\'a}mara et al.(2019)]{LopezCamaraetal2019} L{\'o}pez-C{\'a}mara, D., De Colle, F., \& Moreno M{\'e}ndez, E.\ \textbf{2019}, {\em MNRAS}, 482, 3646

\bibitem[Miszalski et al.(2011)]{Miszalskietal2011} Miszalski, B., Corradi, R.~L.~M., Boffin, H.~M.~J., et al.\ \textbf{2011}, {\em MNRAS}, 413, 1264

\bibitem[Miszalski et al.(2019)]{Miszalskietal2019MNRAS487} Miszalski B., Manick R., Van Winckel H., Miko{\l}ajewska J., \textbf{2019}, {\em MNRAS}, 487, 1040

\bibitem[Mohamed \& Podsiadlowski(2007)]{MohamedPodsiadlowski2007} Mohamed, S., \& Podsiadlowski, P.\ \textbf{2007}, 15th European Workshop on White Dwarfs, 372, 397 

\bibitem[Morris(1987)]{Morris1987} Morris, M.\ \textbf{1987}, {\em PASP}, 99, 1115

\bibitem[Naiman et al.(2020)]{Naimanetal2020} Naiman, B.~V., Sabach, E., Gilkis, A., \& Soker, N.\ \textbf{2020}, {\em MNRAS}, 491, 2736

\bibitem[Nordhaus \& Blackman(2006)]{NordhausBlackman2006} Nordhaus, J., \& Blackman, E.~G.\ \textbf{2006}, {\em MNRAS}, 370, 2004

\bibitem[Orosz et al.(2019)]{Oroszetal2019} Orosz, G., G{\'o}mez, J.~F., Imai, H., et al.\ \textbf{2019}, {\em MNRAS}, 482, L40

\bibitem[Prialnik \& Livio(1985)]{PrialnikLivio1985} Prialnik, D., \& Livio, M.\ \textbf{1985}, {\em MNRAS}, 216, 37

\bibitem[Rechy-Garc{\'\i}a et al.(2020)]{RechyGarciaetal2020} Rechy-Garc{\'\i}a, J.~S., Guerrero, M.~A., Duarte Puertas, S., Chu, Y. -H.; Toala, J. A., \& Miranda, L. F.\ \textbf{2020}, {\em MNRAS}, 492, 1957
 
\bibitem[Rechy-Garc{\'{\i}}a et al.(2017)]{RechyGarciaetal2017}  Rechy-Garc{\'{\i}}a, J., Vel{\'a}zquez, P.~F., Pe{\~n}a, M., \& Raga, A.~C.\ \textbf{2017}, {\em MNRAS}, 464, 2318

\bibitem[Reyes-Ruiz \& L{\'o}pez(1999)]{ReyesRuizLopez1999} Reyes-Ruiz, M., \& L{\'o}pez, J.~A.\ \textbf{1999}, {\em ApJ}, 524, 952

\bibitem[Sabach \& Soker(2018)]{SabachSoker2018} Sabach, E., \& Soker, N.\ \textbf{2018}, {\em MNRAS}, 479, 2249

\bibitem[Sahai et al.(2008)]{Sahaietal2008} Sahai, R., Claussen, M., S{\'a}nchez Contreras, C., Morris, M., \& Sarkar G.\ \textbf{2008}, {\em ApJ}, 680, 483

\bibitem[Sahai \& Trauger(1998)]{SahaiTrauger1998} Sahai, R., \& Trauger, J.~T.\ \textbf{1998}, {\em AJ}, 116, 1357

\bibitem[Santander-Garc{\'\i}a et al.(2019)]{SantanderGarciaetal2019} Santander-Garc{\'\i}a, M., Jones, D., Alcolea, J., Wesson R., \& Bujarrabal V.,\ \textbf{2019}, {\em Highlights on Spanish Astrophysics X}, 392

\bibitem[Schreier et al.(2019)]{Schreieretal2019} Schreier, R., Hillel, S., \& Soker, N.\ \textbf{2019}, {\em MNARS}, 490, 4748

\bibitem[Shiber et al.(2019)]{Shiberetal2019} Shiber, S., Iaconi, R., De Marco, O., et al.\ \textbf{2019}, {\em MNRAS}, 488, 5615

\bibitem[Shiber et al.(2017)]{Shiberetal2017} Shiber, S., Kashi, A., \& Soker, N.\ \textbf{2017}, {\em MNRAS}, 465, L54

\bibitem[Shiber et al.(2016)]{Shiberetal2016} Shiber, S., Schreier, R., \& Soker, N.\ \textbf{2016}, {\em Research in Astronomy and Astrophysics}, 16, 117

\bibitem[Shiber \& Soker(2018)]{ShiberSoker2018} Shiber, S., \& Soker, N.\ \textbf{2018}, {\em MNRAS}, 477, 2584

\bibitem[Soker(1990)]{Soker1990AJ} Soker, N.\ \textbf{1990}, {\em AJ}, 99, 1869

\bibitem[Soker(1992)]{Soker1992} Soker, N.\ \textbf{1992}, {\em ApJ}, 389, 628

\bibitem[Soker(1996)]{Soker1996} Soker, N.\ \textbf{1996}, {\em ApJ}, 468, 774

\bibitem[Soker(2004)]{Soker2004NewA} Soker, N.\ \textbf{2004}, {\em New Astronomy}, 9, 399


\bibitem[Soker(2015)]{Soker2015GEE} Soker, N.\ \textbf{2015}, {\em ApJ}, 800, 114

\bibitem[Soker(2016a)]{Soker2016GEEILOT} Soker, N.\ \textbf{2016a}, {\em New Astronomy}, 47, 16

\bibitem[Soker(2016b)]{Soker2016Rev} Soker, N.\ \textbf{2016b}, 6, {\em New Astronomy Rev.}, 75, 1

\bibitem[Soker(2019)]{Soker2019Termination} Soker, N.\ \textbf{2019}, {\em MNRAS}, 483, 5020

\bibitem[Soker(2020)]{Soker2020jetILOT} Soker, N.\ \textbf{2020}, arXiv:2001.07879

\bibitem[Soker et al.(2019)]{Sokeretal2019CEJSN} Soker, N., Grichener, A., \& Gilkis, A.\ \textbf{2019}, {\em MNRAS}, 484, 4972

\bibitem[Soker \& Rappaport(2001)]{SokerRappaport2001} Soker, N., \& Rappaport, S.\ \textbf{2001}, {\em ApJ}, 557, 256

\bibitem[Tafoya et al.(2019)]{Tafoyaetal2019} Tafoya D., Orosz G., Vlemmings W.~H.~T., Sahai R., P{\'e}rez-S{\'a}nchez A.~F., \textbf{2019}, {\em A\&A}, 629, A8

\bibitem[Tocknell et al.(2014)]{Tocknelletal2014} Tocknell, J., De Marco, O., \& Wardle, M.\ \textbf{2014}, {\em MNRAS}, 439, 2014

\bibitem[Wesson et al.(2018)]{Wessonetal2018}  Wesson R., Jones D., Garc{\'\i}a-Rojas J., Boffin H.~M.~J., Corradi R.~L.~M., \textbf{2018}, {\em MNRAS}, 480, 4589

\bibitem[Witt et al.(2009)]{Wittetal2009} Witt, A.~N., Vijh, U.~P., Hobbs, L.~M., Aufdenberg, J. P., Thorburn, J. A., \& York, D. G.\ \textbf{2009}, {\em ApJ}, 693, 1946

\bibitem[Zou et al.(2020)]{Zouetal2020} Zou, Y., Frank, A., Chen, Z., et al.\ \textbf{2020}, arXiv:1912.01647

\end{thebibliography}
\end{document}